\documentclass[showpacs,prb,superscriptaddress,twocolumn]{revtex4-1}

\usepackage{textcomp}
\usepackage{amsmath}
\usepackage{amssymb}
\usepackage{graphicx} 
\usepackage[lofdepth,lotdepth,caption=false]{subfig}
\usepackage[breaklinks=true,colorlinks,citecolor=blue,linkcolor=blue,urlcolor=blue]{hyperref}

\usepackage{bm}
\usepackage{color}
\usepackage{ulem}
\usepackage{textcomp}

\newcommand{\e}{e}

\renewcommand{\Im}{\mathop{\text{Im}}\nolimits}

\newcommand{\Tr}{\mathop{\text{Tr}}\nolimits}

\newcommand{\us}{\uparrow}
\newcommand{\ds}{\downarrow}

\begin{document}

\title{Andreev levels spectroscopy of topological three-terminal junctions}

\author{Stefano Valentini}
\affiliation{NEST, Scuola Normale Superiore and Istituto Nanoscienze-CNR, I-56127 Pisa, Italy}
\email{stefano.valentini@sns.it}

\author{Rosario Fazio}
\affiliation{NEST, Scuola Normale Superiore and Istituto Nanoscienze-CNR, I-56127 Pisa, Italy}

\author{Fabio Taddei}
\affiliation{NEST, Istituto Nanoscienze-CNR and Scuola Normale Superiore, I-56127 Pisa, Italy}

\begin{abstract}
We calculate the differential conductance at a probe inserted in the weak link of a topological Josephson junction, consisting of a semiconducting nanowire deposited on top of two separated superconductors.
Our aim is to understand how the peculiar features in the
spectrum of Andreev bound states, arising due to the presence of Majorana bound states at the ends of the two topological superconducting wires defining the junction, can be determined through a measurement of the differential conductance. We find that when the probe allows a single propagating mode, the differential conductance presents a dip at zero voltage of zero conductance close to the position where the spectrum exhibits the topologically protected crossing. This can be viewed as a signature of the presence of Majorana states, which does not require fermion parity conservation and is robust against parameters' changes, as well as disorder. On the contrary, when the probe allows two or more propagating modes the differential conductance resembles the spectrum of Andreev bound states. This has been established making use of both numerical and analytical methods.
\end{abstract}

\maketitle

\section{Introduction}
By finding real solutions to the Dirac equation, Majorana predicted the existence of fermions which are their own 
antiparticles.~\cite{Majorana1937} Although never observed as elementary particles, such Majorana fermions can 
actually exist in condensed matter in the form of exotic excitations.~\cite{Kitaev2001} The search for these Majorana modes
in condensed matter systems has been attracting a vast interest primarily because of their topological nature, which 
would allow a protected way of manipulating quantum information.~\cite{Nayak2008} Most of the implementations are 
related to the realization of an effective $p$-wave superconductivity.~\cite{Kitaev2001,Alicea2012,Beenakker2013,Leijnse2012}

The implementation we are focusing on in this work is related to one-dimensional (1D) $p$-wave superconductors and makes use of a 
semiconducting wire, in the presence of spin-orbit interaction, Zeeman fields, and a superconducting order parameter induced 
by an s-wave superconductor located in proximity of the nanowire.
In a certain range of parameters the nanowire is predicted to be topologically non-trivial, exhibiting a pair of Majorana 
bound states (MBS) at its ends.~\cite{Lutchyn2010,Oreg2010} Among the several different signatures of MBS proposed so far, 
an incontrovertible evidence of their existence should derive from the measurement of 
(i) a quantized differential conductance peak at zero voltage, predicted to appear when the nanowire is contacted to a normal 
electrode~\cite{Law2009,Flensberg2010,Wimmer2011}; and (ii) a fractional Josephson effect which should appear when two 
nanowires are coupled through a weak link.~\cite{Kitaev2001,Kwon2004}
In the latter case the Josephson current is predicted to be 4$\pi$-periodic in the superconductors phase difference $\phi$ if the 
parity of the number of fermions in the system is conserved.

Topological superconducting nanowires have been recently experimentally realized by a few groups (see 
Refs.~\onlinecite{Mourik2012,Das2012,Deng2012,Finck2013}). In all of those experiments a differential conductance peak at zero 
voltage, although not quantized, has been observed to persist over a quite wide range of magnetic field values, in accordance with
theoretical expectations. Although there is a  widespread belief that MBS are associated to the anomalies observed in the 
experiments, there is still a vivid debate since a number of spurious effects can give rise to similar zero-voltage conductance peaks. 
On the other hand, $4\pi$-periodicity of the Josephson current has never been observed for such a system.

All the experiments mentioned above~\cite{Mourik2012,Das2012,Deng2012,Finck2013} are conducted in a two-terminal geometry.
A multi-terminal setup will certainly lead to 
a deeper understanding of transport in topological superconductors.~\cite{Weithofer2013}
In this paper we consider a Josephson junction consisting of two 
topological superconducting nanowires separated by a normal weak link in which an additional probe electrode is inserted.
Our work is inspired, although not directly connected, by a recent experiment,~\cite{Chang2013} where the tunneling conductance 
through a segment of a InAs wire confined between two superconducting leads has been measured using a third metallic tunnel probe.
Multi-terminal Josephson devices based on InAs wires connected to several normal and superconducting leads have been also experimentally studied in Refs.~\onlinecite{Roddaro2011,Spathis2011,Giazotto2011}.
The setup we consider is sketched in Fig.\ref{setup}(a). The number of propagating modes reaching the weak link can be effectively 
controlled by means of a quantum point contact (QPC) acting on the probe close to the interface with the weak link.
Our aim is to understand how the presence of Majorana end modes affects the differential conductance $G$ at the probe electrode.
More specifically, we shall assess under which conditions the local density of states (LDOS) in the weak link, determined by the spectrum 
of Andreev bound states (ABS), can be deduced from a measurement of $G$.
Note that the Josephson current and the ABS spectrum in the weak link are  tightly linked. In a topological junction, in particular, 
the $4\pi$-periodicity of the Josephson current is due to ABS which present a topologically protect crossing at $\phi=\pi$. Indeed, in 
the case of Josephson tunnel junction between two $p$-wave superconducting  wires the ABS take the simple form 
$E_{\rm{ABS}}=\pm\Delta_p\sqrt{T}\cos (\phi/2)$, where the sign $\pm$ depends on the fermion parity,  $\Delta_p$ is the gap of the 
superconductors and $T$ is the transmission probability of the tunnel junction~\cite{Kwon2004}.

By analyzing the problem both numerically and analytically, we find that when the probe allows a single propagating mode, the differential 
conductance $G$ at the probe does not reflect the LDOS in the weak link calculated in the absence of the probe. More precisely, while the 
LDOS for $\phi\simeq\pi$ presents a peak with a maximum at zero energy, $G$ develops a zero-voltage dip of zero conductance, whose 
origin can be attributed to perfectly destructing interference effects.
This finding is in agreement with the results of Ref.~\onlinecite{Ioselevich2013}, where the tunneling conductance of discrete Andreev bound 
states has been studied.
Zero conductance at zero voltage has also been found in a network of MBS.~\cite{Flensberg2010,Weithofer2013,Zazunov2013}
Furthermore, we find that the zero-voltage, zero-conductance dip is robust (``topologically protected")
against changes of the parameters, such as spatial asymmetry in the structure and presence of disorder, and can be viewed as a novel signature 
of the presence of Majorana fermions. On the contrary, when the QPC allows two or more propagating modes in the weak link, the interference 
effects giving rise to the dip are largely suppressed and the differential conductance resembles the LDOS in the absence of the probe.

The paper is organized as follows: in Sec.~\ref{model} we detail the system under investigation and the model used to describe it. In 
Sec.~\ref{res} we report the main numerical results obtained for the LDOS (Sec.~\ref{subsec::LDOS}) and for the differential 
conductance (Sec.~\ref{subsec::DCn}). The analytical model 
used to calculate the differential conductance can be found in Sec.~\ref{subsec::G}, while in Sec.~\ref{conc} we summarize and give 
the concluding remarks.

\section{The Model}
\label{model}
\begin{figure}[htb]
\centering
     \includegraphics[width=0.4\textwidth]{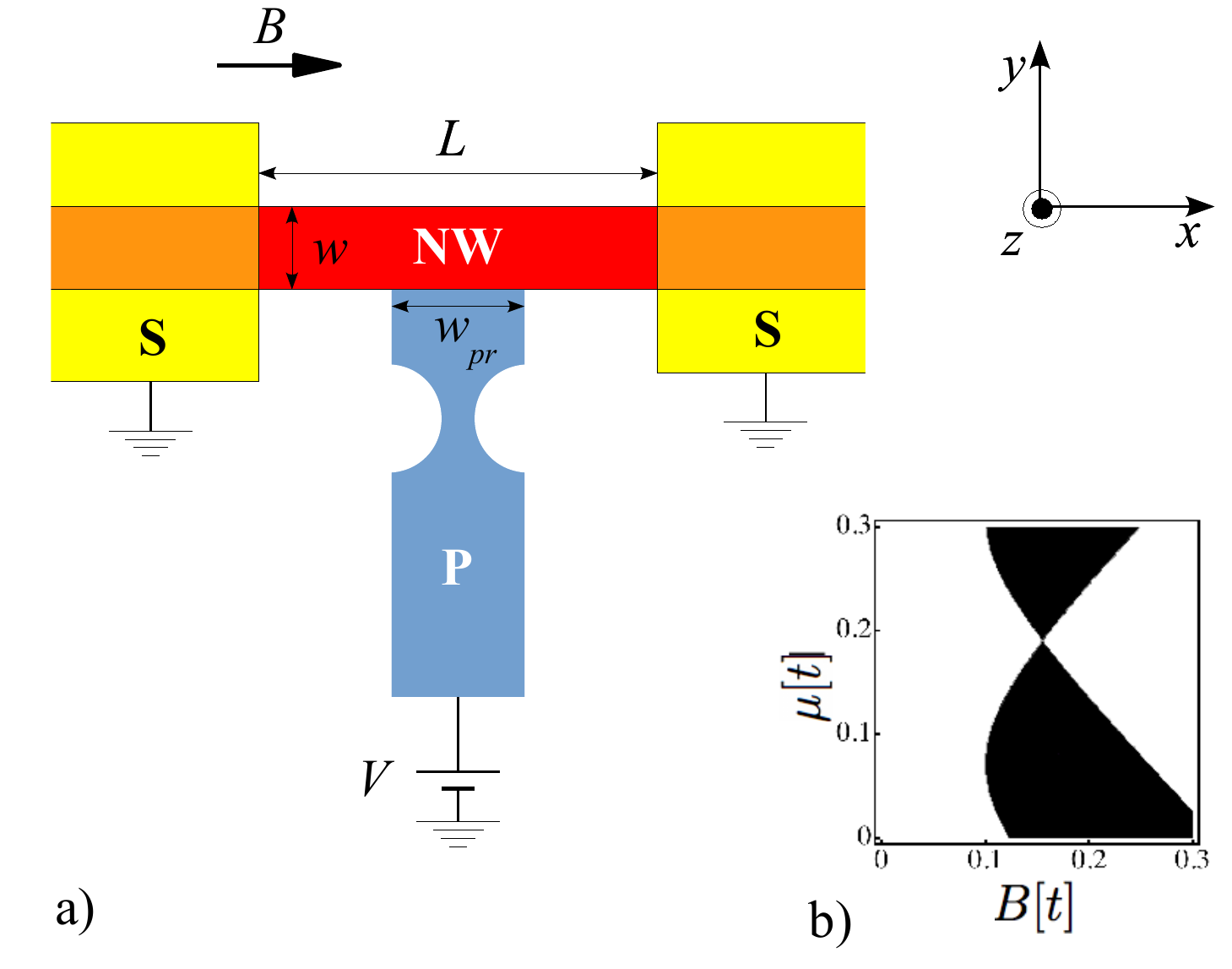}
    \caption{(Color online) (a) Sketch of the system. A semiconducting nanowire (NW) of width $w$ with strong spin-orbit coupling is placed on top 
    of two $s$-wave superconducting electrodes (S). A magnetic field is applied parallel to the axis of the nanowire ($x$-direction). A non-vanishing 
    superconducting pairing is induced, through the proximity effect, in nanowire sectors in contact to the superconductors (orange/lighter regions). 
    A normal probe (P) of width $w_{pr}$ is attached to the non-proximized sector of the nanowire, whose length is $L$. The width of the 
    probe is reduced (quantum point contact) in the vicinity of its interface with the nanowire. A voltage bias $V$ is applied to the probe, while the 
    superconductors are grounded. (b) Phase diagram of an ideal, isolated topological superconducting nanowire as a function of the Zeeman 
    field $B$ and of the chemical potential $\mu$, in units of $t$. The width of the nanowire is assumed to be $w=10 a$, where $a$
    is the lattice constant.
    The non-trivial topological phase is marked in black.}
    \label{setup}
\end{figure}

The system, sketched in Fig.~\ref{setup} (a), is composed of a semiconducting nanowire with strong spin-orbit coupling placed on top of 
two $s$-wave superconductors, which differ by the phase of the order parameter. Through the proximity effect a non-vanishing superconducting 
pairing is induced in those regions of the nanowire that are in contact to the superconductors, thus realizing a 
superconductor-normal-superconductor (SNS) Josephson junction. A magnetic field is applied parallel to the axis of the nanowire  ($x$-direction), 
which is attached to a normal probe in its N region -- the weak link.

The two-dimensional tight-binding Hamiltonian of the nanowire, of width $w$, reads
\begin{eqnarray}\label{eq:hamiltnw}
\hat {\cal H}_{\rm NW} &=& -t\sum_{\langle i,j\rangle,\sigma}\hat c^\dag_{i,\sigma}\hat c_{j,\sigma} + (\varepsilon_0-\mu)\sum_{ i,\sigma}\hat 
c^\dag_{i,\sigma}\hat c_{i,\sigma} \nonumber\\
&& + i\alpha \sum_{\langle i,j\rangle,\sigma,\sigma'} (\nu'_{ij}\sigma^x_{\sigma\sigma'} - \nu_{ij}\sigma^y_{\sigma\sigma'})\hat 
c^\dag_{i,\sigma}\hat c_{j,\sigma'} \nonumber\\
&&  + B \sum_{i,\sigma,\sigma'}\sigma^x_{\sigma\sigma'}\hat c^\dag_{i,\sigma}\hat c_{i,\sigma'} \;,
\end{eqnarray}
where $\hat c^\dag_{i,\sigma}$ is a creation operator for an electron of spin $\sigma$ in site $i$, and the symbol $\langle i,j\rangle$ stands for nearest-neighbor sites $i$ and $j$.
Here $t$ is the hopping energy, $\varepsilon_0= 4 t$
is a uniform on-site energy which sets the zero of energy,  $\alpha$ is the Rashba spin-orbit (SO) 
coupling strength, $B$ is the Zeeman field along the wire, $\sigma^{i}$ are spin-1/2 Pauli matrices, $\nu_{i j} = \hat{\bm x} \cdot \hat {\bm d}_{i j}$, and 
$\nu'_{i j} = \hat{\bm y} \cdot \hat {\bm d}_{i j}$ with $\hat {\bm d}_{i j} = ({\bm r}_i - {\bm r}_j)/|{\bm r}_i - {\bm r}_j|$ being the unit vector connecting 
site $j$ to site $i$.

The probe is characterized by the same parameters as the nanowire, but without SO coupling, and its Hamiltonian reads
\begin{eqnarray}\label{eq:hamiltpr}
\hat {\cal H}_{\rm P} &=& -t\sum_{\langle i,j\rangle,\sigma}\hat c^\dag_{i,\sigma}\hat c_{j,\sigma} + 
\sum_{ i,\sigma} (\varepsilon_0-\mu+U_i) \hat c^\dag_{i,\sigma}\hat c_{i,\sigma} \nonumber\\
&&  + B \sum_{i,\sigma,\sigma'}\sigma^x_{\sigma\sigma'}\hat c^\dag_{i,\sigma}\hat c_{i,\sigma'}~,
\end{eqnarray}
where $U_i$ is a position-dependent potential which is non-zero in the presence of a QPC (see below).
The coupling to the nanowire is accounted for by the following term
\begin{eqnarray}\label{eq:hamiltc}
\hat {\cal H}_{\rm C} &=& -t\gamma_{pr} \sum_{\langle i,j\rangle,\sigma}^{(\text{I})}\hat c^\dag_{i,\sigma}\hat c_{j,\sigma} +  \text{H.c.} \;,
\end{eqnarray}
where the superscript (I) in the sum indicates that  the sites $i$ and $j$ are at the interface between the probe ($i$) and the nanowire ($j$),
while $\gamma_{pr}$ is the strength of the coupling. 
Finally the induced pairing in the nanowire is accounted for by the term:
\begin{equation}\label{eq:SChamilt}
\hat {\cal H}_{\rm S} = \sum_{ i}\left[\Delta_i~\hat c^\dag_{i,\us}\hat c^\dag_{i,\ds} + \text{H.c.}\right]~.
\end{equation}
The superconducting order parameter $\Delta_i$ is assumed piecewise constant, with $|\Delta_i|=\Delta$ in the regions in contact with the superconductor and $\Delta_i=0$ in the central region of length $L$.
The phases of the superconductors on the left and on the right with respect to this region differ by $\phi$.
Moreover, we assume that a barrier can be present at the boundary between the S and N regions of the nanowire, leading to a decrease of
the value of the hopping energy $t$ and of the SO coupling $\alpha$ by a factor $\gamma_{L/R}$ for the left and right boundary, respectively.The complete Hamiltonian of the system then reads
\begin{equation}\label{eq:FULLhamilt}
\hat {\cal H} = \hat {\cal H}_{\rm NW} + \hat{\cal H}_{\rm S} + \hat {\cal H}_{\rm P} + \hat {\cal H}_{\rm C}~.
\end{equation}

The proximized regions of the nanowire are in the topological non-trivial phase, thus hosting pairs of Majorana bound states at their ends, for appropriate 
values of the parameters. For an ideal, isolated single-channel wire, the topological non-trivial phase occurs when $B>\sqrt{\mu^2+\Delta^2}$.
~\cite{Lutchyn2010,Oreg2010} For a nanowire of finite width, the phase diagram presents a richer structure~\cite{Luchyn2011,Stanescu2011,
Potter2011,marco} where trivial and non-trivial topological regions alternate [see, as an example, Fig.~\ref{setup}(b)].

\section{Results}
\label{res}
Numerical calculations are performed within the tight-binding model sketched above using a recursive Green's function technique. \cite{Sanvito1999}
The parameters are chosen such that only one channel is open in the nanowire: For the topological phase we choose $\mu = 0$ and $B=0.2 t$, 
while for the trivial phase we choose $\mu=0.15 t$ and $B=0.1 t$. In both cases $w=10 a$ ($a$ being the lattice constant), $\Delta=0.1 t$, and $\alpha=0.1 t$.
The LDOS is calculated in the weak link in the absence of a probe. The differential conductance $G=dI/dV$ is calculated, using a wave-function-matching 
technique, \cite{Zwierzycki2008} for a probe of width $w_{pr}$ attached to the nanowire in a given position, where $I$ is the current flowing through 
the probe and $V$ is the bias voltage applied to it (the superconductors are grounded). The coupling to the probe is controlled by the coupling parameter 
$\gamma_{pr}$ [so that the hopping energy at the probe-wire interface is $t\gamma_{pr}$ 
as indicated in Eq.~(\ref{eq:hamiltc})] and by the presence of a QPC, which restricts the effective 
number of open channels of the probe in the contact region.
The QPC is described by the following saddle-like potential \cite{buttiker}:
\begin{equation}
\label{QPC}
U(x,y) = \mbox{max} \Big\{ 0 ,  U_0 - U_y (\frac{y + d}{a})^2 + U_x (\frac{x - w_{pr}/2}{a})^2 \Big\},
\end{equation}
where $d$ is the distance of the saddle point from the probe-wire interface. $U_0$, $U_x$ and $U_y$ are external parameters that fix the shape of the potential and are related to the characteristics of the gate used to realize the QPC. 
The condition $U_x \gg U_y$ must be satisfied in order to clearly see quantized conductance.\cite{buttiker}

\subsection{Local density of states in the absence of a probe} \label{subsec::LDOS}
We anticipate that the results of this section are in complete agreement with the ones present in the literature (see, for example, Refs.~\onlinecite{Lutchyn2010,Black-Schaffer2011,Tewari2012,marco}).
We will use them as a reference for the differential conductance results shown in the next section.
The LDOS can be computed through the relation
\begin{equation}
\label{ldos_green}
{\cal N}({\bm r},E) = -\frac{1}{2 \pi}\Im\{{\rm Tr}[{\cal G}({\bm r},E)]\}~,
\end{equation}
where ${\cal G}({\bm r},E)$ is the Green's function and the factor 2 in the denominator is introduced to avoid a double counting of particle and hole degrees-of-freedom intrinsic in the formalism we use.
We shall consider both the short ($L\ll\xi$) and the long ($L\gg\xi$) junction regimes, where $\xi=\hbar v_F/E_g$ is the superconducting coherence length ($v_F$ being the Fermi velocity and $E_g$ being the induced gap in the energy spectrum of the NW). Within the tight-binding model $\xi\simeq(k_Fa)ta/E_g$, where $k_F$ is the Fermi momentum.

\begin{figure}[tb]
    \centering
     \subfloat[][]{\label{LDOSst}
     \includegraphics[width=0.22\textwidth]{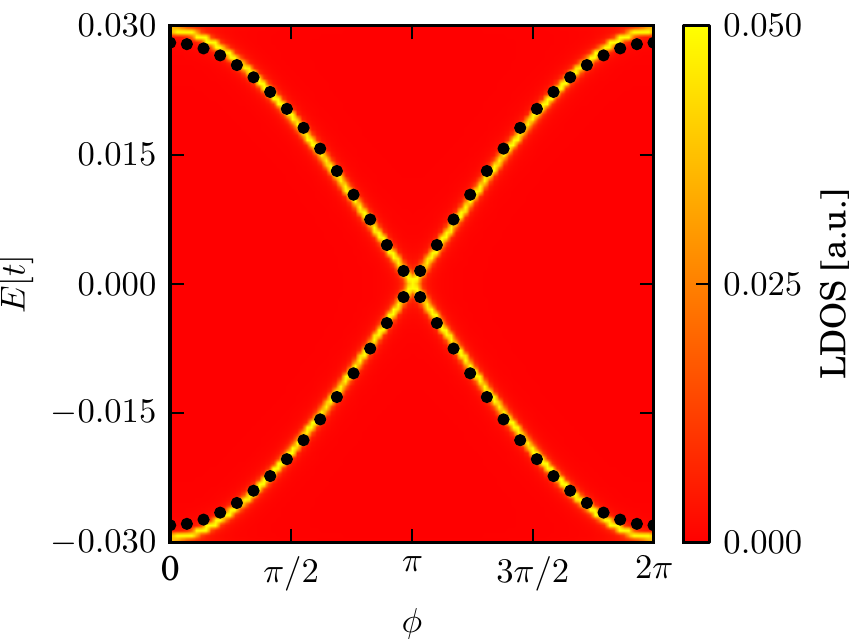}}
     \subfloat[][]{\label{LDOSsr}
     \includegraphics[width=0.22\textwidth]{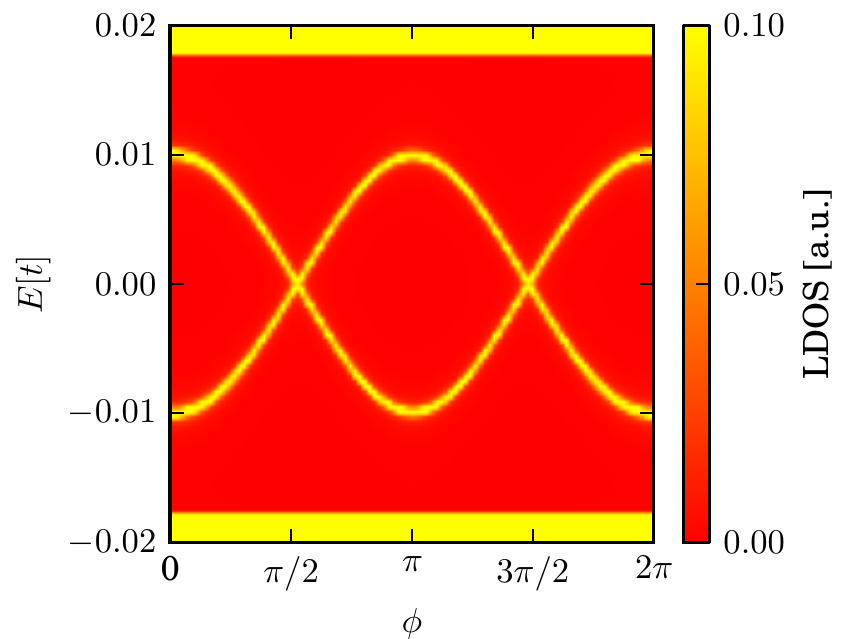}}\\
      \subfloat[][]{\label{LDOSlt}
     \includegraphics[width=0.22\textwidth]{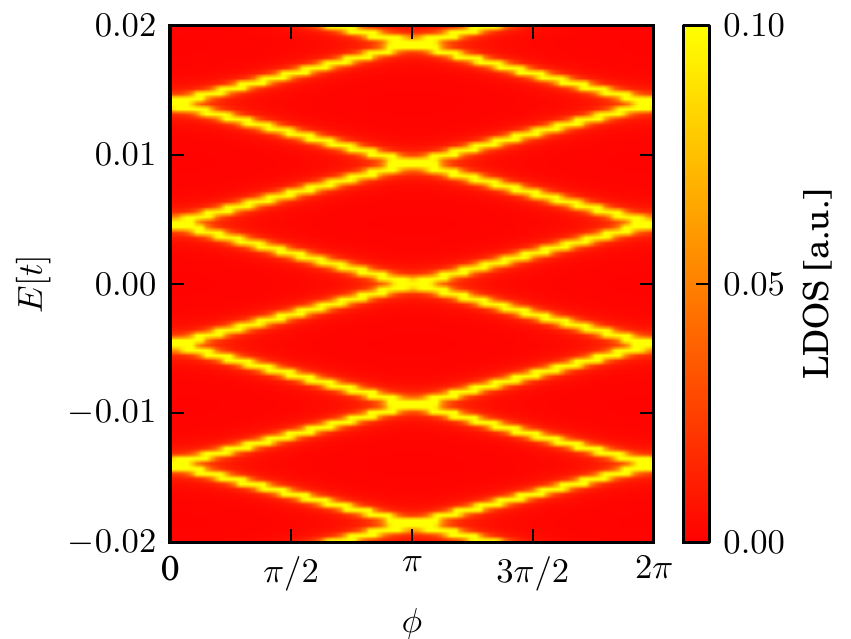}}
       \subfloat[][]{\label{LDOSlr}
     \includegraphics[width=0.22\textwidth]{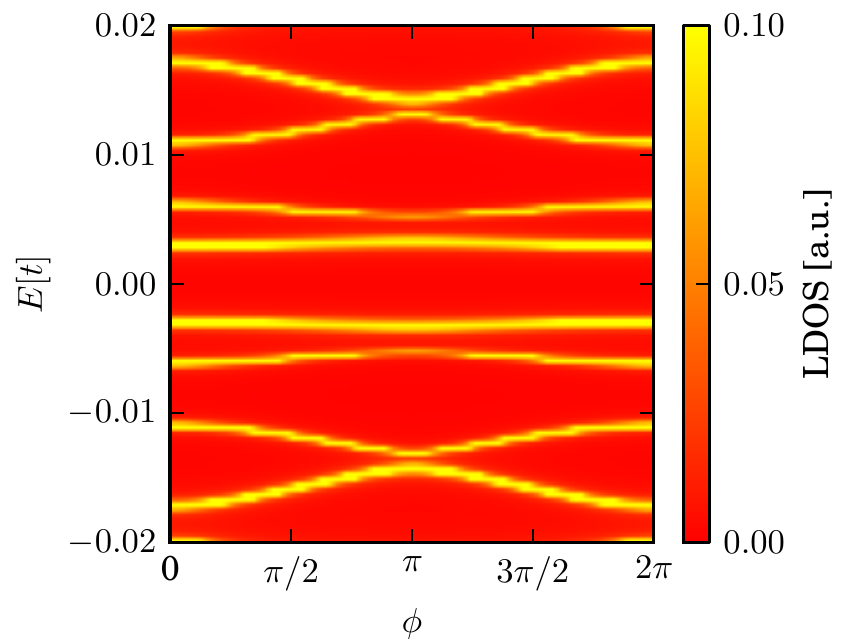}}
    {\caption{(Color online) Density plots of the LDOS as a function of  phase difference $\phi$ and energy $E$ at a site in the middle of the normal region assuming $\gamma_{L/R} = 1$. Note that, in order to increase the contrast, we have plotted the LDOS in a restricted range of values.
    Plots (\ref{LDOSst}), topological phase ($B = 0.2 t, \mu = 0$), and (\ref{LDOSsr}), trivial phase ($B = 0.1 t, \mu = 0.15 t$), refer to a short junction ($L = 2 a$). Plots (\ref{LDOSlt}), topological phase, and (\ref{LDOSlr}), trivial phase, refer to a long junction ($L = 200 a$).
The ABS spectrum in plot (\ref{LDOSst}) is in good accordance (black dots) with the expression $E_{\text{ABS}}=\pm\Delta_p \sqrt{T}\cos (\phi/2)$, relative to a $p$-wave Josephson junction, setting $T=1$ and $\Delta_p=0.028 t$.
}
    \label{LDOS}}
\end{figure}

\begin{figure}[htb]
    \centering
     \subfloat[][]{\label{DOSnsST}
     \includegraphics[width=0.2\textwidth]{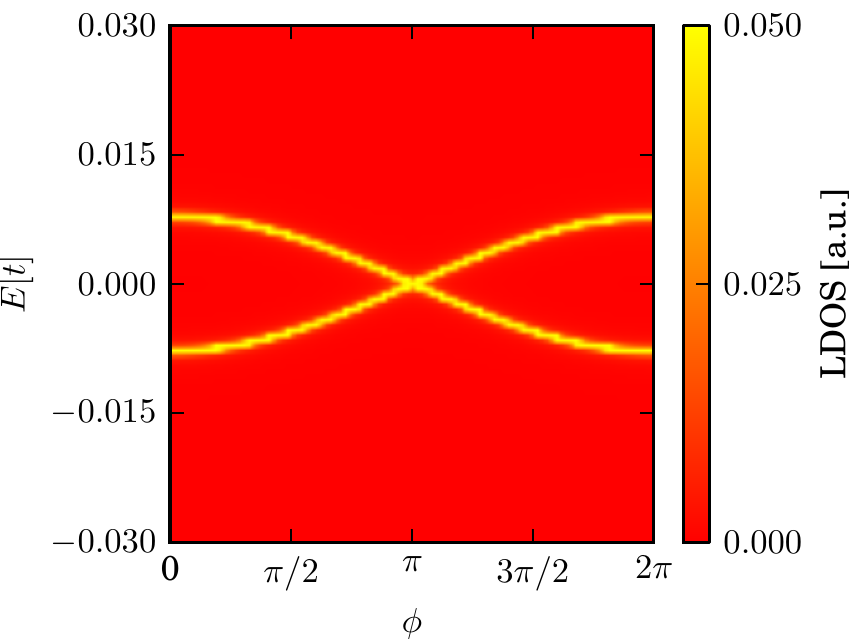}}
     \subfloat[][]{\label{DOSnsSR}
     \includegraphics[width=0.2\textwidth]{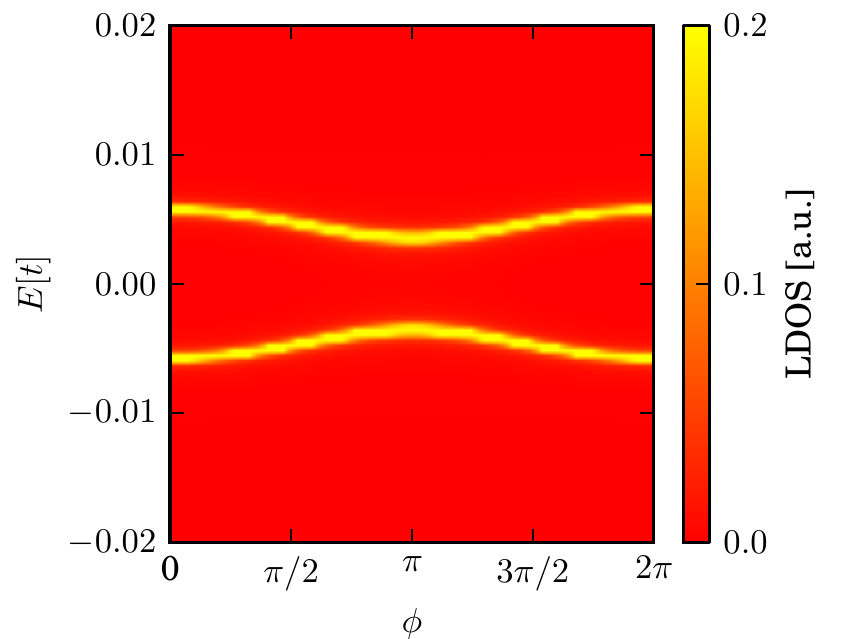}}\\
      \subfloat[][]{\label{DOSnsLT}
     \includegraphics[width=0.2\textwidth]{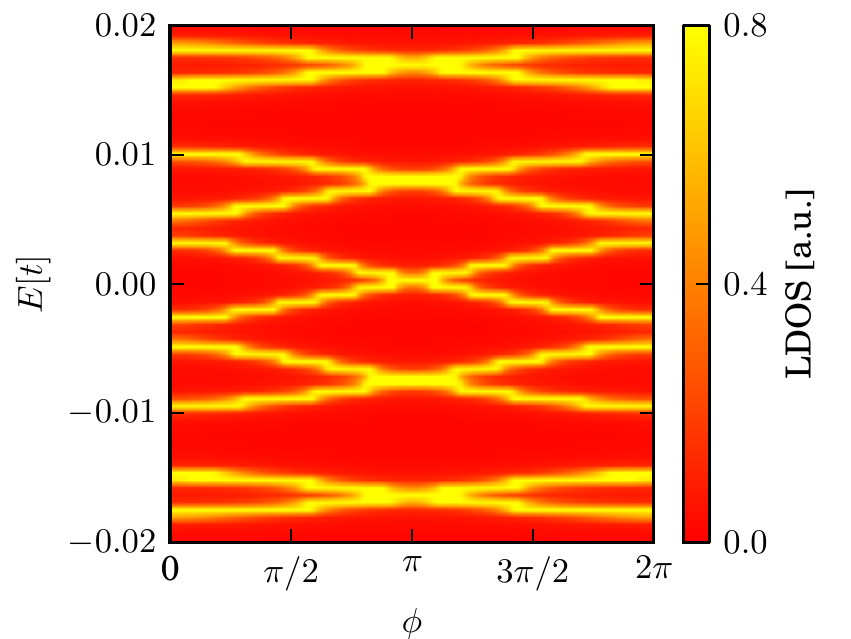}}
       \subfloat[][]{\label{DOSnsLR}
     \includegraphics[width=0.2\textwidth]{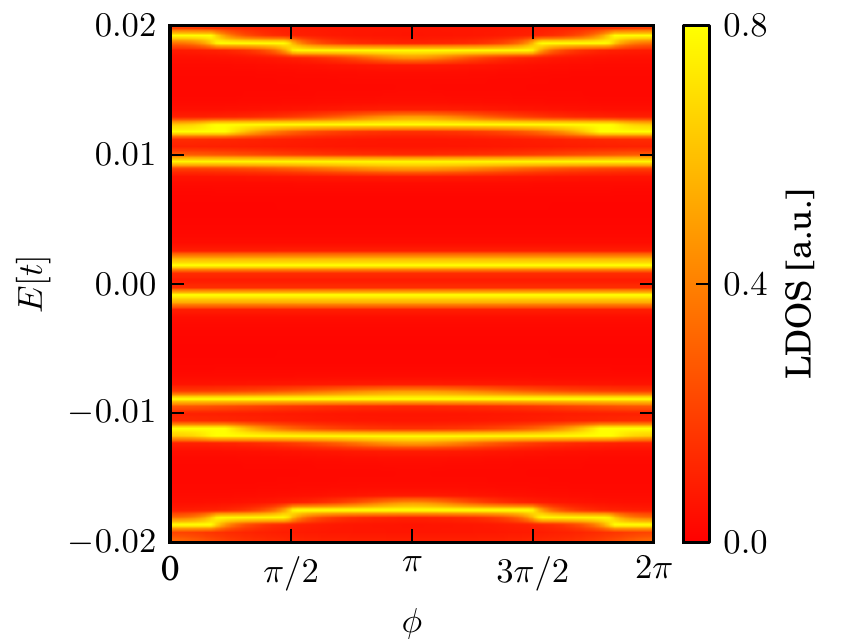}}
    {\caption{The same as for Fig.~\ref{LDOS}, but setting $\gamma_{L/R} = 0.7$.}
    \label{DOSbarrNS}}
\end{figure}

In Fig.~\ref{LDOS} the LDOS, relative to a site in the middle of the weak link, is plotted as a function of energy $E$ and phase difference $\phi$ (the LDOS is qualitatively equal over the whole normal region).
Figures~\ref{LDOSst} and~\ref{LDOSsr} refer to a short junction, while Figs.~\ref{LDOSlt} and~\ref{LDOSlr} refer to a long junction.
In all the plots, ABS manifest themselves as sharp peaks of large amplitude (lighter color).
For a short junction ($L=2a$) in the topological phase (Fig. \ref{LDOSst}) the spectrum of ABS is characterized by a crossing at $E=0$ and $\phi=\pi$ and it is in good accordance with the prediction for a 1D Josephson junction between spinless $p$-wave superconductors, namely $E \propto \pm \cos(\phi/2)$ -- see the black dots in Fig.~\ref{LDOSst}.
Such crossing is topologically protected, e.g., it persists even when the couplings $\gamma_{L/R}$ are changed.
In the trivial phase [Fig. \ref{LDOSsr}], the spectrum of ABS presents two crossings at zero energy, which are not protected as their positions in $\phi$ change with $\gamma_{L/R}$, 
e.g. see Fig.~\ref{DOSbarrNS}.
In both cases, at fixed phase, we have only two ABS with opposite energy.
On the contrary, for a long junction ($L=200a$) multiple pairs of ABS are present at fixed phase and the dispersion relations acquire a richer structure.
Still, we find one crossing at $E=0$ and $\phi=\pi$ in the topological phase [Fig. \ref{LDOSlt}] and no crossings in the trivial one [Fig. \ref{LDOSlr}].

\begin{figure}[tb]
    \centering
     \includegraphics[width=0.4\textwidth]{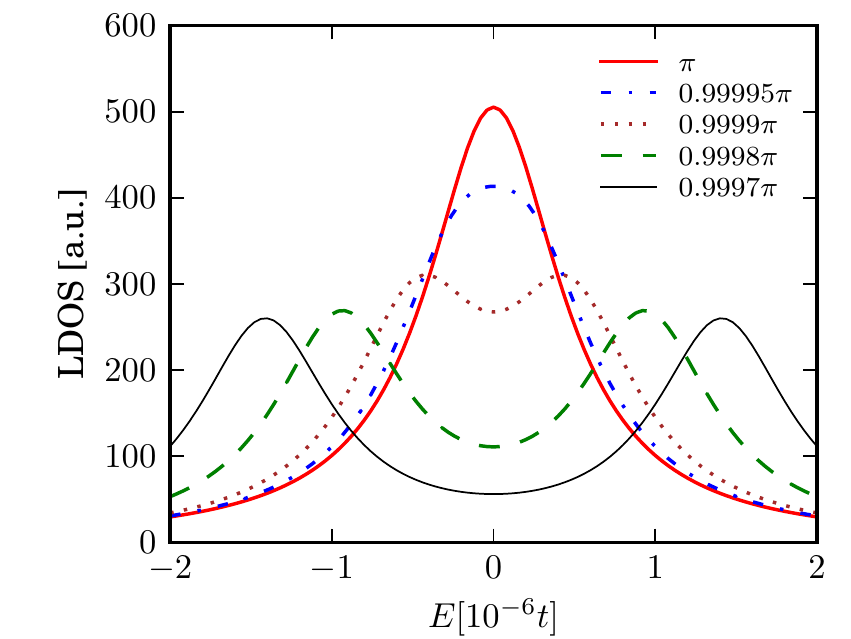}
    {\caption{(Color online) Vertical cuts of Fig.~\ref{LDOSlt} over a reduced range of energies. LDOS as a function of energy $E$ for different values of $\phi$ close to $\pi$, when the junction is in the topological phase. The LDOS is calculated at a site in the middle of the normal region assuming $L=200 a$, $\gamma_{L/R}=1$ and $\eta=10^{-7} t$.}
    \label{ldosfase}}
\end{figure}

In Fig.~\ref{ldosfase} the LDOS of a long junction in the topological phase is plotted as a function of the energy for different values of $\phi$ near $\pi$.
We observe that two ABS-peaks are clearly distinguishable for value of $\phi$ far enough from $\pi$, while they overlap when $\phi$ gets closer to $\pi$ eventually merging into a single peak.
Notice that the overlap is an incoherent sum of the two peaks (for example, at the crossing $\phi = \pi$ the LDOS is twice as high as a single ``isolated" ABS).
A (small) imaginary part $\eta$ has been added to the energy $E$ in Eq.~(\ref{ldos_green}) to carry out the calculation.
A finite $\eta$ (taken to be equal to $10^{-7} t$ in all plots) physically originates from relaxation processes due to the coupling to the environment.
The width (height) of the peaks in the LDOS increases (decreases) with increasing $\eta$.

The effect of a reduced coupling $\gamma_{L/R}$ can be appreciated by comparing Fig.~\ref{LDOS} with Fig.~\ref{DOSbarrNS}, where 
all parameters are the same but $\gamma_{L/R}=0.7$. No major effects emerge in the topological phase apart from 
a reduction of the bandwidth [Fig.~\ref{DOSnsST}], which additionally causes the disappearance of the crossings at $\phi=0$ and 
$2\pi$ [Fig.~\ref{DOSnsLT}].
This is a consequence of the fact  that the crossing at $E=0$ and $\phi=\pi$ in the topological phase is protected against changes of the coupling between 
proximized and normal sectors of the nanowire and of the length of the normal region. 
On the other hand, in the trivial phase the ABS spectra are more strongly modified, in particular for the short 
junction where the crossings at zero energy disappear [Fig.~\ref{DOSnsSR}], but the number of crossings at zero energy is always even.

We are now aiming at comparing the LDOS presented in this section to the differential conductance $G$ calculated at the probe inserted in the weak link.
We shall furthermore assess how $G$ is affected by the presence of Majorana bound states at the boundaries between the proximized regions and the weak link.

\subsection{Differential conductance: Numerical calculations}
\label{subsec::DCn}
When a voltage $V$ is applied to the probe, and $V$ is smaller than the superconducting gap $\Delta$, the transmission through the 
proximized sectors of the nanowires is suppressed and the differential conductance at the probe is determined, at zero temperature, by
\begin{equation}
G(V)= \left. \frac{2 e^2}{h}\Tr[r_a(E)^\dag r_a(E)]\right|_{E=eV},
\end{equation}
where $r_a(E)$ is the Andreev reflection matrix at the interface with  the probe, evaluated at energy $E$.
The matrix $r_a(E)$, whose dimension is determined by the number of open channels in the probe, is numerically calculated, as mentioned in 
Sec.~\ref{model}, through a recursive Green's function method~\cite{Sanvito1999} combined with a wave-function-matching technique. \cite{Zwierzycki2008}
We assume that the  spin-orbit interaction  is absent in the probe. Furthermore, we consider the case of a a weak probe-wire coupling ($\gamma_{pr}=0.1$).
The probe is attached to the nanowire in the middle of the normal region and, initially, it is assumed to be narrow enough ($w_{pr}= 8 a$) 
to support a single open channel (moreover, the QPC is absent).
We checked that when $\gamma_{L}=0$ or $\gamma_{R}=0$, i. e., when a single topological superconductor is present, $G$ shows a quantized 
zero-bias peak (ZBP) signaling the presence of a Majorana bound state.
From now on we shall consider completely transparent interfaces between proximized and normal sectors of the nanowire ($\gamma_{L/R}=1$).

\begin{figure}[tb]
    \centering
     \subfloat[][]{\label{Gst}
     \includegraphics[width=0.2\textwidth]{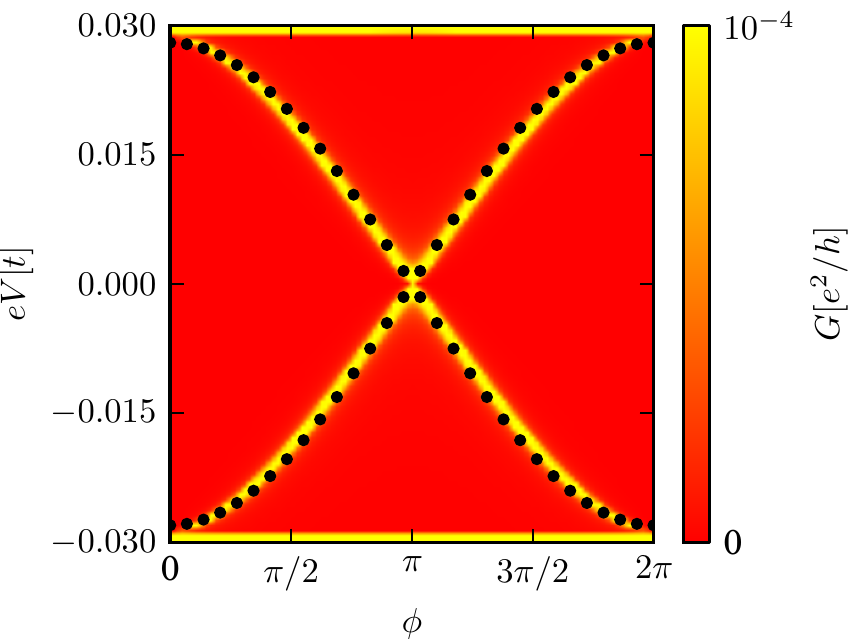}}
     \subfloat[][]{\label{Gsr}
     \includegraphics[width=0.2\textwidth]{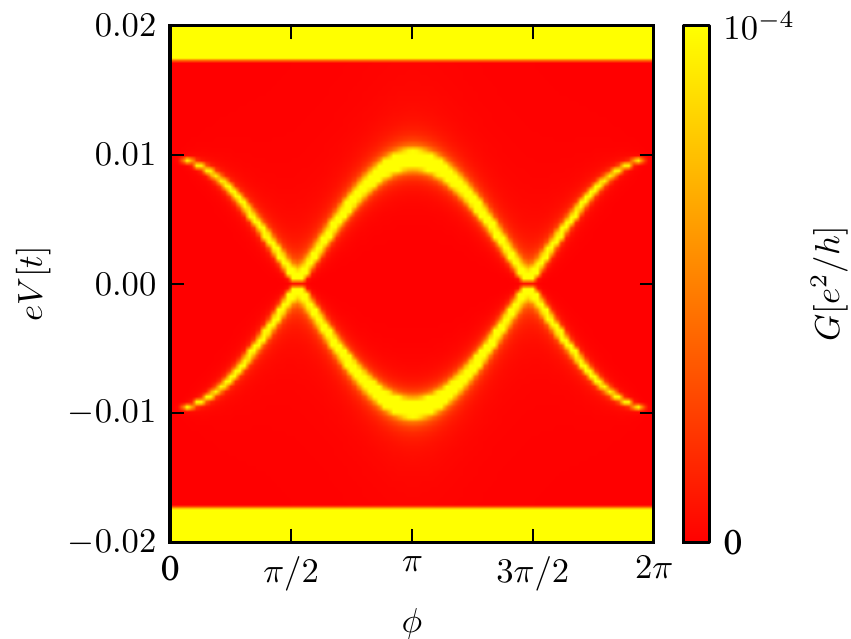}}\\
     \subfloat[][]{\label{Glt}
     \includegraphics[width=0.2\textwidth]{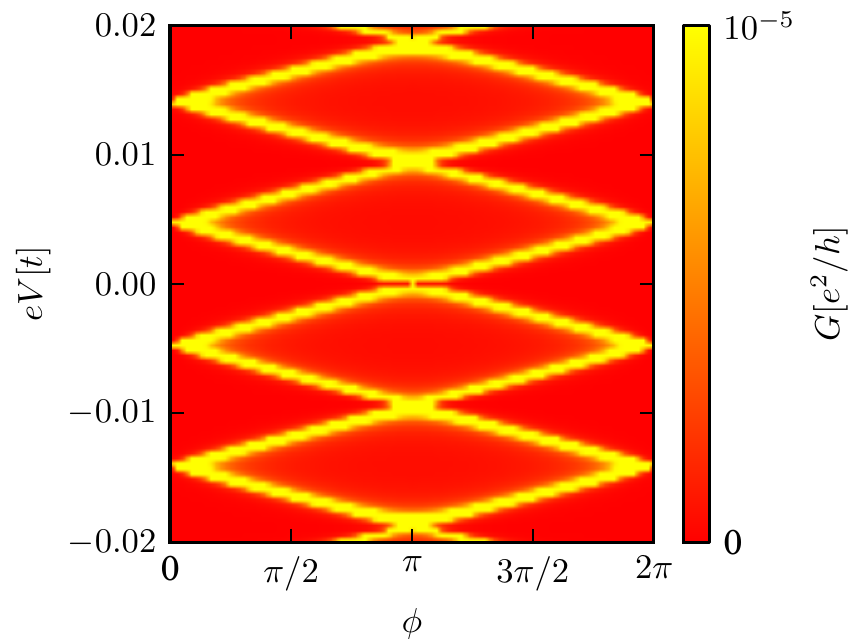}}
     \subfloat[][]{\label{Glr}
     \includegraphics[width=0.2\textwidth]{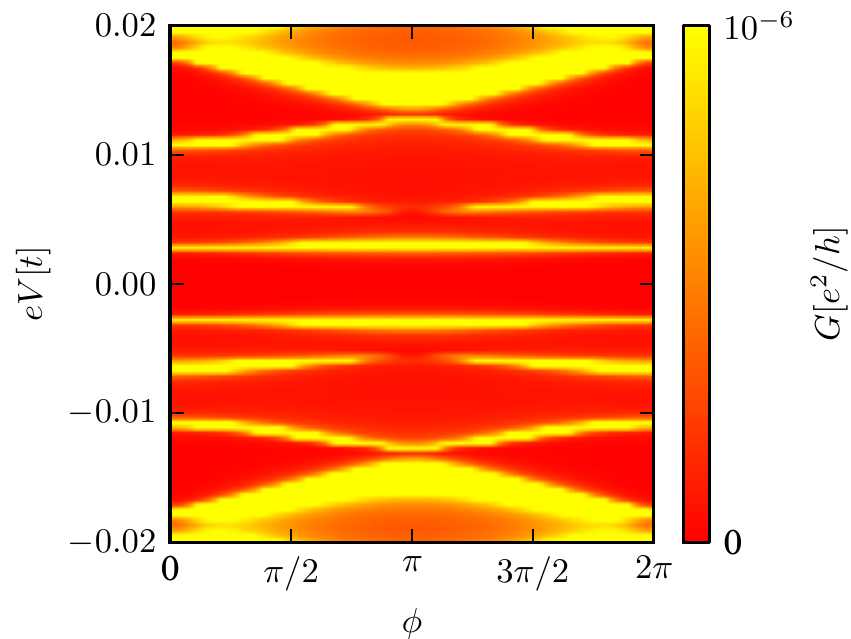}}
       {\caption{(Color online) Density plot of the differential conductance $G$ as a function of phase difference $\phi$ and applied voltage $V$, for $\gamma_{L/R} = 1$.
       The single-channel probe is attached in the middle of the normal region with coupling $\gamma_{pr}=0.1$.
       Note that, in order to increase the contrast, we have plotted $G$ in a restricted range of values.
       Plots (\ref{Gst}) (topological, $B = 0.2 t, \mu = 0$) and (\ref{Gsr}) 
       (trivial, $B = 0.1 t, \mu = 0.15 t$) refer to a short junction ($L=2a$). Plots (\ref{Glt}) (topological) and (\ref{Glr}) (trivial) refer to a long junction ($L = 200 a$). Lighter colors mean 
       higher conductance. The conductance in plot (\ref{Gst}) is in good accordance (black dots) with Eq.~(\ref{fit}) assuming small $\Gamma$ as 
       in Eq.~(\ref{fitlim}) and setting $|\Delta'| k_F=0.028 t$.}
    \label{Gplot}}
\end{figure}
\begin{figure}[tb]
    \centering
     \includegraphics[width=0.4\textwidth]{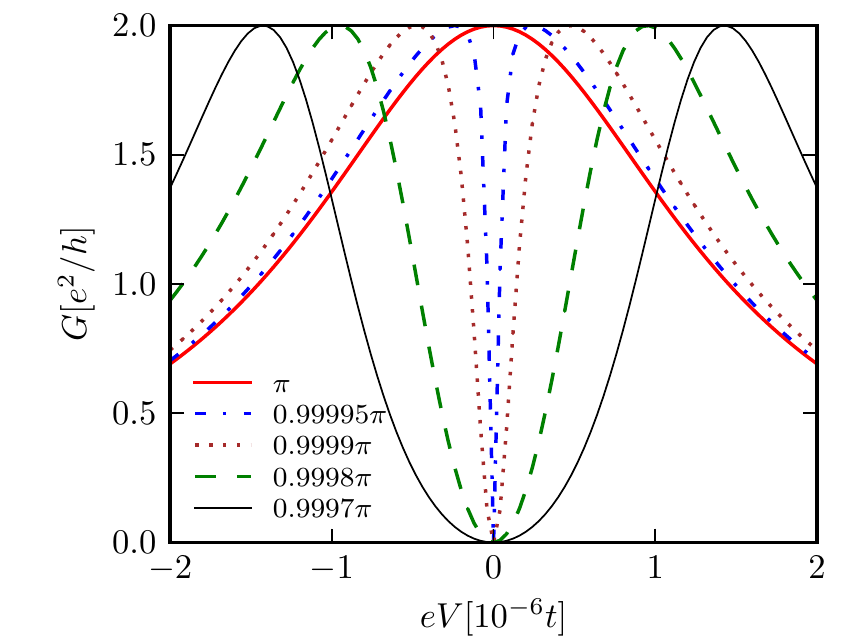}
    {\caption{(Color online) Vertical cuts of Fig.~\ref{Glt} over a reduced range of voltages. Differential conductance $G$ plotted as a function of applied voltage $V$ for values of $\phi$ 
    close to $\pi$ and $\gamma_{L/R} = 1$, when the junction is in the topological phase. The single-channel probe is attached in the middle of the normal 
    region with coupling $\gamma_{pr}=0.1$ and $L=200a$.}
    \label{Gfase}}
\end{figure}

In Fig.~\ref{Gplot}, $G$ is plotted as a function of bias voltage $V$ and phase difference $\phi$ for both short and long junctions in the topological and 
trivial phases (the panels are organized as in Fig.~\ref{LDOS}).
At first glance, one notices that Fig.~\ref{Gplot} remarkably resembles the LDOS plots of Fig.~\ref{LDOS}.
Namely, $G$ exhibits sharp peaks of height $2e^2/h$ in apparent correspondence to the ABS.
A more careful look, however, shows that this is not true in the crucial region around zero energy and close to $\phi=\pi$.
This can be appreciated by analyzing the data for the topological phase close to such a crucial region.
In Fig.~\ref{Gfase} the differential conductance for a long junction ($L=200a$) is plotted as a function of $V$ for some values of $\phi$ close to $\pi$.
At $\phi=\pi$ a quantized peak, in units of $2e^2/h$, develops centered in $V=0$.
If $\phi$ is not exactly equal to $\pi$, however, $G$ exhibits a zero-conductance dip at zero voltage, whose width increases by moving away from $\pi$.
The two peaks, of quantized height, occur at energies corresponding to ABS peaks in the LDOS (see Fig.~\ref{ldosfase}), but, unlike the LDOS, they do 
add coherently.~\cite{nota_peaks} More precisely, a complete destructive interference occurs at zero voltage (this interpretation is corroborated by 
the analytical results reported in Sec.~\ref{subsec::G}). The occurrence of a zero-conductance dip at zero voltage can be also understood in terms of 
the properties of the scattering matrix in the presence of particle-hole symmetry. According to B\'eri,~\cite{Beri2009} for a single-channel probe, the 
eigenvalues of the Andreev reflection probability matrix are bound, at zero energy, to be either 0 or 1. As a result, at $\phi=\pi$, the presence of a resonant 
level (the topologically protected crossing between the two branches of the ABS) imposes $|r_a(0)|^2=1$. On the other hand, its absence at $\phi\ne\pi$ 
forces $|r_a(0)|^2=0$. For the characterization of peaks and dip in terms of the parameters defining the system we refer the reader to Sec.~\ref{subsec::G}, 
where analytical expressions are provided. Analogous behavior (not shown) is obtained for the conductance of a short junction.

We stress that the zero-conductance dip at zero voltage is topologically robust: It persists for variations of $\gamma_{R/L}$, $\gamma_{pr}$, $L$, and, 
within the topological phase, chemical potential $\mu$ and magnetic field $B$. As we shall see below, it also persists when disorder is introduced in the 
normal region of the nanowire.

Let us finally consider the trivial phase.
Nothing significant happens for a long junction, where no crossings occur at zero energy [conductance and LDOS plots are virtually coinciding, see 
Figs.~\ref{LDOSlr} and \ref{Glr}, respectively].
For a short junction, however, the perturbation induced by the probe gives rise to two barely noticeable anti-crossings at zero energy [see Fig.~\ref{Gsr}].

\begin{figure}[tb]
    \centering
     \includegraphics[width=0.4\textwidth]{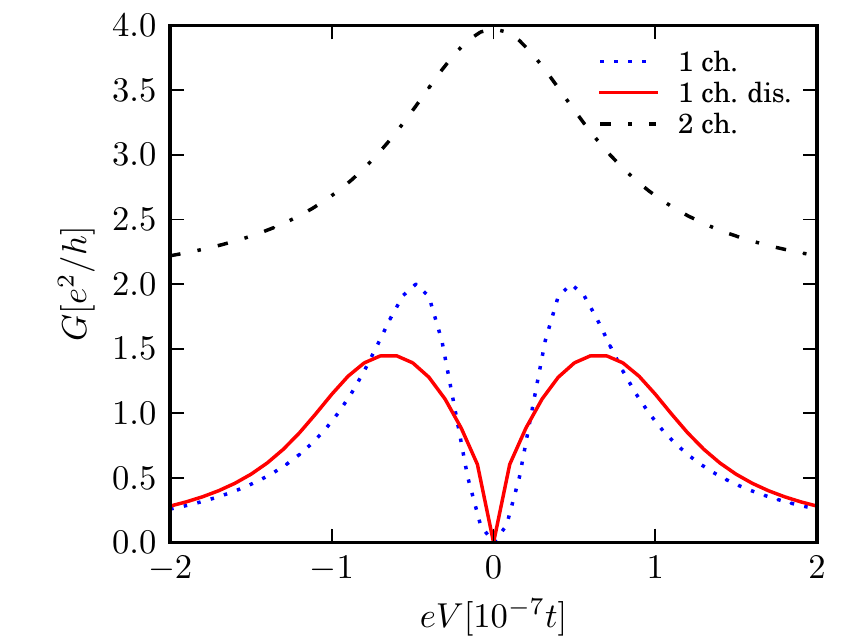}
    {\caption{(Color online) Differential conductance when the probe includes a QPC. $G$ is plotted as a function of applied voltage $V$ at 
    $\phi=0.99999 \pi$ and $\gamma_{L/R} = 1$, for a long junction ($L=200a$) in the topological phase. The blue dotted line refers to the 
    case where the QPC allows a single channel ($U_0 = 0.1 t$), while the black dashed-dotted line refers to the double-channel case ($U_0=0$). 
    The red full line is relative to a single-channel QPC in the presence of disorder, described as a random on-site energy in the range $[-0.001 t, 0.001 t]$, 
    and we average over 100 configurations. The probe, with QPC at distance $d=150 a$, is attached in the middle of the normal region with coupling 
    $\gamma_{pr}=0.1$. Other parameters are: $U_x= 0.00005 t$, $U_y = 0.005 t$, and $w_{pr} = 21 a$.}
    \label{qpc099999}}
\end{figure}

We now relax the condition of a single channel in the probe and consider the case of two open channels.
This can be simply done by increasing the probe width $w_{pr}$, without changing other parameters. 
\begin{figure}[b]
    \centering
     \includegraphics[width=0.4\textwidth]{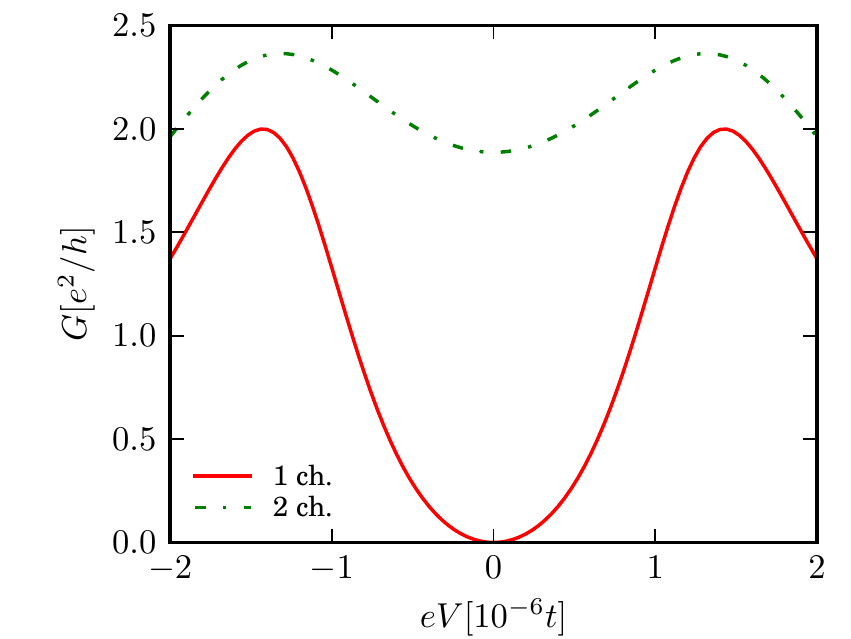}
    {\caption{(Color online) Differential conductance $G$ plotted as a function of applied voltage $V$ at $\phi=0.9997 \pi$ and $\gamma_{L/R} = 1$, 
    for a long junction ($L=200a$) in the topological phase. The probe is attached in the middle of the normal region with coupling $\gamma_{pr}=0.1$. 
    The red full line refers to a single-channel probe of width $w_{pr}=8 a$, while the green dashed-dotted line refers to a double-channel probe of width $w_{pr}=14 a$.}
    \label{G2ch}}
\end{figure}
The resulting differential conductance is plotted in Fig.~\ref{G2ch} for a long junction, with $\phi$ close to $\pi$.
Contrary to the single-channel case (red full line), for two channels (green dashed-dotted line) the zero-conductance dip at $V=0$ is not present and
 $G$ resembles more accurately the LDOS (compare with the black full line in Fig. \ref{ldosfase}).
More realistically, the effective number of propagating modes reaching the weak link can be controlled through a QPC inserted in the probe (see Fig.~\ref{setup}).
By varying $U_0$, in Eq.~(\ref{QPC}), one can change the number of channels allowed by the QPC without changing the geometry of the system.
Of course, the constriction reduces the effective strength of the coupling between probe and wire, so the width of the conductance peaks decreases. 
As a consequence, the qualitative behavior of the conductance with a narrow probe is the same as the one for a wider probe with a QPC, but on a 
reduced energy scale. This is evident from Fig.~\ref{qpc099999}, where $G$ is plotted as a function of voltage, with a QPC allowing two channels 
(black daseh-dotted line) and one channel (blue dotted line).
We notice that, for the two-channel case, the single peak shown in the plot is a result of the ``incoherent'' sum of two peaks having a large overlap.

To complete the numerical study, we have investigated how disorder affects the zero-conductance dip in the topological phase.
Disorder is described as a random on-site energy introduced both in the weak link and in the region of the probe where the QPC is located.
For a QPC with a single open channel it turns out that the peaks (of quantized height) are randomly shifted in voltage for each configuration of disorder, 
but the differential conductance at $V=0$ is always zero. As a consequence, the averaging over 100 disorder configurations leads to lower and broader 
peaks (see red full line in Fig.~\ref{qpc099999}), but  still with a zero-conductance dip.

\subsection{Differential conductance: Analytical model} \label{subsec::G}
Several features found in the numerical simulations can be understood with a simple model amenable to an analytical solution.
In this section we calculate the differential conductance  for the same setup considered so far where both the nanowire and the probe are
replaced by strictly one-dimensional wires.
This analytical calculation allows us to single out the main features found in the numerical calculations.
The system consists of two semi-infinite spinless $p$-wave superconducting nanowires coupled through a normal region (the weak link), which is connected to 
a normal spinless semi-infinite probe.
The Hamiltonian for the 1D $p$-wave superconductors in the Bogoliubov-De Gennes formalism is~\cite{Pientka2013}
\begin{equation}\label{pwavehamiltonian}
H= \begin{pmatrix}
\frac{p^2}{2 m}-\mu  &  \frac{1}{2}\left\{ \Delta', p \right\}\\
\frac{1}{2}\left\{ \Delta'^*, p \right\}  &  -\frac{p^2}{2 m}+\mu\\
\end{pmatrix},
\end{equation}
where $\Delta'$ is the $p$-wave superconducting coupling and $\mu$ is the chemical potential. The Hamiltonian for the probe and the weak link 
is the same as Eq.~(\ref{pwavehamiltonian}) after taking $\Delta'=0$.
The weak-link/probe junction is described by a three-leg beam splitter, whose scattering matrix reads:
\begin{equation}
\label{fork}
S_B=\left(
\begin{array}{ccc}
  \sqrt{1-2\Gamma} & \sqrt{\Gamma} & \sqrt{\Gamma} \\
\sqrt{\Gamma}& a & b \\
\sqrt{\Gamma}& b & a\\
\end{array}
\right) ,
\end{equation}
where $a=1/2(1-\sqrt{1-2\Gamma})$, $b=1/2(-1-\sqrt{1-2\Gamma})$,
with $0\le\Gamma\le 1/2$ being the transmission of the probe/nanowire interface.
This can be seen as an effective model for the structure described in Sec. \ref{model} with a single-channel probe and the topological 
superconducting nanowire deep in the topological phase (i.e., far away from the gap closure).

The Andreev reflection amplitude at the probe is determined by composing the scattering matrices of the beam splitter [Eq.~(\ref{fork})], of the 
free propagations in the normal regions and of the NS interfaces.\cite{Taddei2004}
The latter have been calculated by matching the wave functions at the two sides of an interface and imposing current conservation.
In the case of completely transparent NS interfaces and for a short junction, we obtain the following expression for the differential conductance:
\begin{widetext}
\begin{equation}\label{Gequ}
G(V)=\frac{2e^2}{h}\frac{\Gamma^2 (eV)^2 |\Delta'|^2 k_F^2 \sin^2(\frac{\phi}{2})}{\Gamma^2 (eV)^2 |\Delta'|^2 k_F^2 \sin^2(\frac{\phi}{2})+
[(eV)^2 \sqrt{1-2\Gamma} - |\Delta'|^2 m \mu (1-\Gamma+\sqrt{1-2\Gamma}) \cos^2(\frac{\phi}{2})]^2},
\end{equation}
\end{widetext}
where $k_F=\sqrt{2 m \mu}$ is the Fermi momentum.

Equation~(\ref{Gequ}) nicely reproduces the numerical results of Sec.~\ref{subsec::G} for a single-channel probe and, in particular, describes 
well the dip at $V=0$ [see the black dots in Fig.~\ref{Gst}]. 
First of all, we notice that, if $\phi\ne\pi$, $G$ goes to zero quadratically in the limit $V \rightarrow 0$.
However, for $\phi=\pi$ the differential conductance at $V=0$ is quantized to its maximum $G(0)=2 e^2/h$.
More generally, one finds that $G$ takes its maximum value ($G=2 e^2/h$) for 
\begin{equation}\label{fit}
(eV)^2 = \frac{|\Delta'|^2 k_F^2 (1-\Gamma+\sqrt{1-2\Gamma}) \cos^2(\frac{\phi}{2})}{2 \sqrt{1-2\Gamma} }.
\end{equation}

In order to obtain an analytic characterization of the features of the differential conductance, let us now assume a small coupling between wire 
and probe (tunneling regime $\Gamma\rightarrow 0$).
In this limit, the following equality holds
\begin{equation} \label{fitlim}
\frac{1-\Gamma+\sqrt{1-2\Gamma}}{2 \sqrt{1-2\Gamma} }=1+ O(\Gamma^2),
\end{equation}
and the differential conductance presents two maxima at $eV=\pm |\Delta'| k_F |\cos(\frac{\phi}{2})|$ [the separation between the peaks' top 
amounts to $\delta_{\text{peak}}=2|\Delta'| k_F |\cos(\frac{\phi}{2})|$].
This means that the position of the peaks does not depend on probe/wire coupling, as long as the latter is weak, and exactly matches the energy of the ABS.
Moreover, the full width of each peak at half maximum (FWHW) is $\sigma_{\text{peak}}=\Gamma k_F |\Delta'| |\sin \frac{\phi}{2}|$, thus the peaks 
broaden with increasing $\Gamma$ and approaching $\phi=\pi$.
We can now consider separately the two situations: (i) phase far from $\pi$ $(|\cot \frac{\phi}{2}| \gg \Gamma/2)$, and (ii) phase close to 
$\pi$ $(|\cot \frac{\phi}{2}| \ll \Gamma/2)$. In case (i) the two conductance peaks are located far apart, i.e., $\sigma_{\text{peak}}\ll\delta_{\text{peak}}$, 
and each peak looks almost symmetrical with respect to its maximum.
In case (ii) the two conductance peaks are close to each other, i.e., $\sigma_{\text{peak}}\gg\delta_{\text{peak}}$, and each peak is highly asymmetrical 
with respect to its maximum.
Notice that the FWHM of each peak does not change much by varying the phase, being $\sin \frac{\phi}{2}\sim 1$.
Moreover, we can characterize the dip through its width, defined as the separation of the two peaks at half maximum $\sigma_{\text{dip}}=
2 |\Delta'| k_F \frac{\cos^2 (\frac{\phi}{2})}{\Gamma |\sin \frac{\phi}{2}|}$.
We find that $\sigma_{\text{dip}}$ is inversely proportional to $\Gamma$ (it increases by weakening the probe/wire coupling) and that the 
dip width is smaller than the distance between the maxima ($\sigma_{\text{dip}}\ll\delta_{\text{peak}}\ll\sigma_{\text{peak}}$).
All the above results are fully compatible with the numerical results of Sec.~\ref{subsec::G}, even for the long junction case.

So far we have assumed zero temperature.
For finite, small temperatures with respect to the energy gap ($k_\text{B} T\ll |\Delta'|k_F$) we have
\begin{widetext}
\begin{equation}
G(V)=\frac{2e}{h}\int_{-\infty}^{\infty}\frac{\Gamma^2 \epsilon^2 |\Delta'|^2 k_F^2 \sin^2(\frac{\phi}{2})}{\Gamma^2 \epsilon^2 |\Delta'|^2 k_F^2 \sin^2(\frac{\phi}{2})+[\epsilon^2 \sqrt{1-2\Gamma} - |\Delta'|^2 m \mu (1-\Gamma+\sqrt{1-2\Gamma}) \cos^2(\frac{\phi}{2})]^2}\Big(- \frac{d f(T,V,\epsilon)}{d \epsilon}\Big) d\epsilon,
\end{equation}
\end{widetext}
where $f$ is the Fermi distribution function
\begin{math}
f(T,V,\epsilon)=1/(\e^{\frac{\epsilon-eV}{k_\text{B} T}}+1).
\end{math}
Essentially, a finite temperature smoothens the conductance peaks (by broadening them and lowering their maxima) and effectively lowers the resolution of the measurement.
Within the tunneling regime, in order to clearly distinguish the two conductance peaks, at a given phase $\phi$, one must have $k_B T \ll \delta_{\text{peak}}$.
On the other hand, to detect the dip at $\phi$ close to $\pi$ ($|\cot \frac{\phi}{2}| \ll \Gamma/2$) one needs to fulfill the more stringent condition $k_B T \ll\sigma_\text{dip}$.
For example, taking $|\Delta'|k_F=250$ $\mu$eV (see Ref.~\onlinecite{Mourik2012}), $\Gamma=0.2$ and $\phi=0.97\pi$ one gets $\sigma_\text{dip}/k_B\sim 60$ mK.

\begin{figure}[b]
    \centering
     \includegraphics[width=0.4\textwidth]{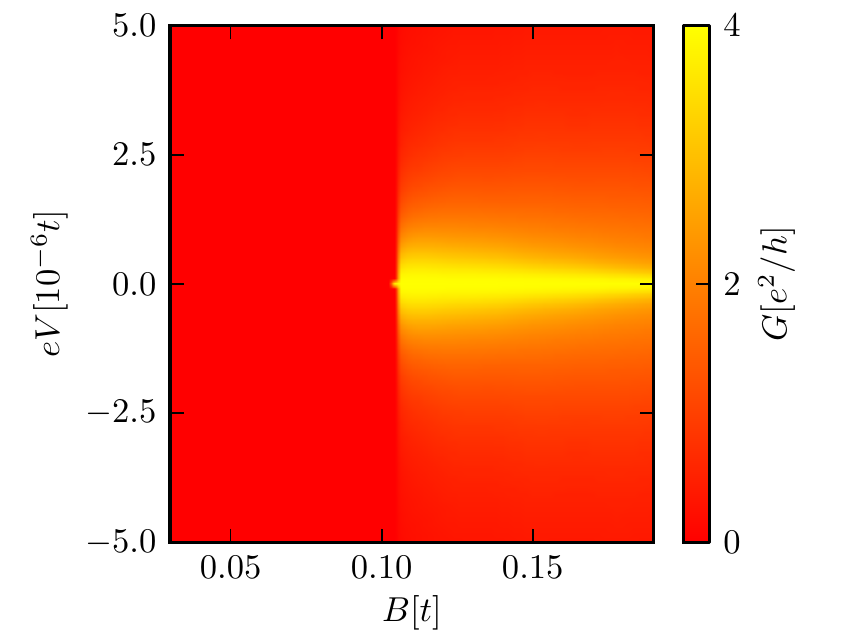}
    {\caption{(Color online) Phase transition. The differential conductance $G$, for $\phi = \pi$, is plotted as a function of Zeeman field $B$ and applied 
    voltage $V$, at $\gamma_{L/R} = 1$. A two-channel probe ($w_{pr}=14a$) is attached in the middle of the normal region with coupling $\gamma_{pr}=0.1$. For this plot we have set $\mu=0.1 t$.}
    \label{phasetransiz}}
\end{figure}
\section{Discussion and Conclusions} \label{conc}
In summary, we have calculated the differential conductance in a three-terminal topological Josephson junction aiming at identifying the effects peculiar of the 
topological phase as compared to the trivial one. The system is based on a semiconducting nanowire with strong spin-orbit coupling and a longitudinally applied Zeeman field.
Superconductivity is induced, through the proximity effect, in the two lateral sectors of the nanowire by placing under them two superconducting electrodes.
For a strong enough Zeeman field, such nanowire sectors access the superconducting topological phase characterized by the appearance of two Majorana bound 
states at the end points of each sector. A Josephson junction is thus formed by leaving a central part uncovered -- the weak link.
The third terminal -- the probe -- is introduced by contacting the weak link to a normal (spin-orbit free) electrode, which is kept at voltage $V$ (the superconductors are grounded).
The differential conductance, related to the current flowing through the third electrode, has been then calculated as a function of the bias voltage $V$ and the superconducting phase difference $\phi$ established between the two superconducting electrodes.
This has been done through numerical calculations based on the tight-binding model and recursive Green's function approach.
The numerical results for the conductance have been reproduced by a 1D analytical model, through which we also studied the impact of a finite temperature.

We have found that a characteristic feature of the Andreev bound states energy spectrum of a topological Josephson junction, namely the topologically protected level 
crossing at zero-energy and $\phi=\pi$ (the origin of the fractional Josephson effect), is reflected in a peculiar way in the differential conductance.
More precisely, in the presence of a single propagating mode in the probe, we find a zero-conductance dip at zero voltage when $\phi$ is close to $\pi$, which is 
topologically robust against changes of the parameters (including disorder and length of the weak link).
This feature, unlike the $4\pi$-periodicity of the Josephson current, does not depend on the fermion parity and can be viewed as a signature of the topological phase.
The zero-conductance dip, however, disappears when more than one propagating mode is allowed in the probe and $G$ resembles  the local density of states 
(i. e. the energy spectrum) of the Josephson junction.

The local density of states in nanostructures is typically measured, with spatial and energy resolution, through the scanning tunneling microscopy.
We notice that the lack of correspondence between differential conductance and local density of states comes with no surprise in our system, where 
the spectrum consists of discrete levels (see Ref.~\onlinecite{Ioselevich2013} for a thorough discussion).
The presence of two channels in the probe, though, is sufficient to restore the $G$ vs. spectrum correspondence.

As shown in Fig.~\ref{ldosfase}, the topological phase is characterized by topologically protected ABSs at $E=0$ and $\phi=\pi$.
This feature can be exploited to observe the phase transition in the differential conductance at the probe. Indeed, for a fixed chemical potential and fixed 
phase $\phi=\pi$, one can observe the emergence of the zero-bias conductance peak by raising the external magnetic field. This can be seen in 
Fig.~\ref{phasetransiz}, where the differential conductance for a two-channel probe, calculated as in Sec.~\ref{subsec::DCn}, is plotted as a function of the Zeeman field $B$ and the bias voltage $V$.
The figure shows that two different regions can be clearly distinguished in the plot and a threshold magnetic field for the onset of the 
topological phase can be easily estimated.
In particular, for $B$ smaller than the threshold, $G$ is virtually zero in the considered voltage range -- trivial phase. On the other hand, a zero bias 
conductance peak is present over the threshold and that region corresponds to the topological phase.   

We conclude by mentioning that a single-channel probe can be realistically realized by placing a QPC in the probe in the vicinity of the interface with the wire.
Although challenging, the zero-conductance dip at zero voltage could be experimentally observed with InAs or InSb nanowires for low enough temperatures.

\section*{Acknowledgments}
We would like to thank V. Giovannetti who participated in this work at the early stages.
We furthermore acknowledge fruitful discussions with K. Flensberg and M. Gibertini.
This work has been supported by the EU FP7 Programme ("Simulations and Interfaces with Quantum Systems" - Grant Agreement  
No. 600645-SIQS, Grant Agreement  
No. 234970-NANOCTM, and Grant Agreement No. 248629-SOLID), by  MIUR-PRIN "Collective quantum phenomena: From strongly correlated systems to quantum simulators" and by the MIUR-FIRB-IDEAS project RBID08B3FM.


\begin{thebibliography}{10}
\bibitem{nota_peaks}
	Note that the position of the LDOS and conductance peaks, at a given value of $\phi$, slightly differs. While the incoherent sum of a pair of peaks tends to bring the peaks closer to each other (LDOS), the destructive interference between the two peaks in the conductance tends to move them away.
\end{thebibliography}
\end{document}